\documentclass[a4paper,11pt]{article}
\pdfoutput=1 

\usepackage{jcappub} 

\usepackage[T1]{fontenc} 

\usepackage{float}
\usepackage{caption}
\usepackage{subfig}

\usepackage[export]{adjustbox}
\usepackage[dvipsnames]{xcolor}
\newcommand{\be}{\begin{equation}}
\newcommand{\ee}{\end{equation}}
\newcommand{\bea}{\begin{eqnarray}}
\newcommand{\eea}{\end{eqnarray}}
\newcommand{\beann}{\begin{eqnarray*}}
\newcommand{\eeann}{\end{eqnarray*}}
\newcommand{\nn}{\nonumber}

\title{\boldmath Echoes from the Abyss: A highly spinning black hole remnant for the binary neutron star merger GW170817}

\author[a,b,c,d,1]{Jahed Abedi,\note{Corresponding author.}}
\author[e,f,g]{Niayesh Afshordi}

\affiliation[a]{Max-Planck-Institut for Gravitationsphysik (Albert-Einstein-Institut), 
D-30167 Hannover, Germany}
\affiliation[b]{Leibniz Universität Hannover, 
D-30167 Hannover, Germany}
\affiliation[c]{Department of Physics, Sharif University of Technology, 
P.O. Box 11155-9161, Tehran, Iran}
\affiliation[d]{School of Particles and Accelerators, Institute for Research in Fundamental Sciences (IPM), 
 P.O. Box 19395-5531, Tehran, Iran}
 \affiliation[e]{Department of Physics and Astronomy, University of Waterloo, 
Waterloo, ON, N2L 3G1, Canada}
\affiliation[f]{Waterloo Centre for Astrophysics, University of Waterloo, 
Waterloo, ON, N2L 3G1, Canada}
\affiliation[g]{Perimeter Institute for Theoretical Physics, 
31 Caroline St. N., Waterloo, ON, N2L 2Y5, Canada}

\emailAdd{jahed.abedi@aei.mpg.de}
\emailAdd{nafshordi@pitp.ca}

\abstract{The first direct observation of a binary neutron star (BNS) merger was a watershed moment in multi-messenger astronomy.  However, gravitational waves from GW170817 have only been observed prior to the BNS merger, but electromagnetic observations all follow the merger event. While post-merger gravitational wave signal in general relativity is too faint (given current detector sensitivities), here we present the first tentative detection of post-merger gravitational wave ``echoes'' from a highly spinning ``black hole'' remnant. The echoes may be expected in different models of quantum black holes that replace event horizons by exotic Planck-scale structure and tentative evidence for them has been found in binary black hole merger events.  The fact that the echo frequency is suppressed by $\log M$ (in Planck units) puts it squarely in the LIGO sensitivity window, allowing us to build an optimal model-agnostic search strategy via cross-correlating the two detectors in frequency/time. We find a tentative detection of echoes at $f_{\rm echo} \simeq 72$ Hz, around 1.0 sec after the BNS merger, consistent with a 2.6-2.7 $M_\odot$ ``black hole'' remnant with dimensionless spin $0.84-0.87$. Accounting for all the ``look-elsewhere'' effects, we find a significance of $4.2 \sigma$, or a false alarm probability of $1.6\times 10^{-5}$, i.e. a similar cross-correlation within the expected frequency/time window after the merger cannot be found more than 4 times in 3 days. If confirmed, this finding will have significant consequences for both physics of quantum black holes and astrophysics of binary neutron star mergers.}

\begin{document}
\maketitle
\flushbottom

\section{Introduction}\label{Introduction}
One of the most controversial questions in quantum gravity today is whether quantum effects may become significant outside horizons of black holes \cite{Braunstein:2009my,Almheiri:2012rt,Maldacena:2013xja,Giddings:2017mym}, or that they only play a role well inside the horizon (e.g., \cite{Abedi:2015yga}).  While the latter is the conservative assumption, the former might come about through (so-far speculative) processes that could resolve the {\it information paradox} \cite{Almheiri:2012rt,Lunin:2001jy,Lunin:2002qf,Maldacena:2013xja}, or the {\it cosmological constant problem(s)} \cite{PrescodWeinstein:2009mp}. In order to resolve the paradox, significant deviations form classicality need to occur by the Page time $\sim M^3$, but they may already be in place as early as the {\it scrambling time} $\sim M \log M$ (or more precisely $T^{-1}_{\rm H} \log S_{\rm BH}$, for temperature $T_{\rm H}$ and entropy $S_{\rm BH}$) \cite{Hayden:2007cs, Sekino:2008he}, where $M$ is the black hole mass in Planck units. 

\begin{figure}[b]
\centering
\includegraphics[width=0.7\textwidth]{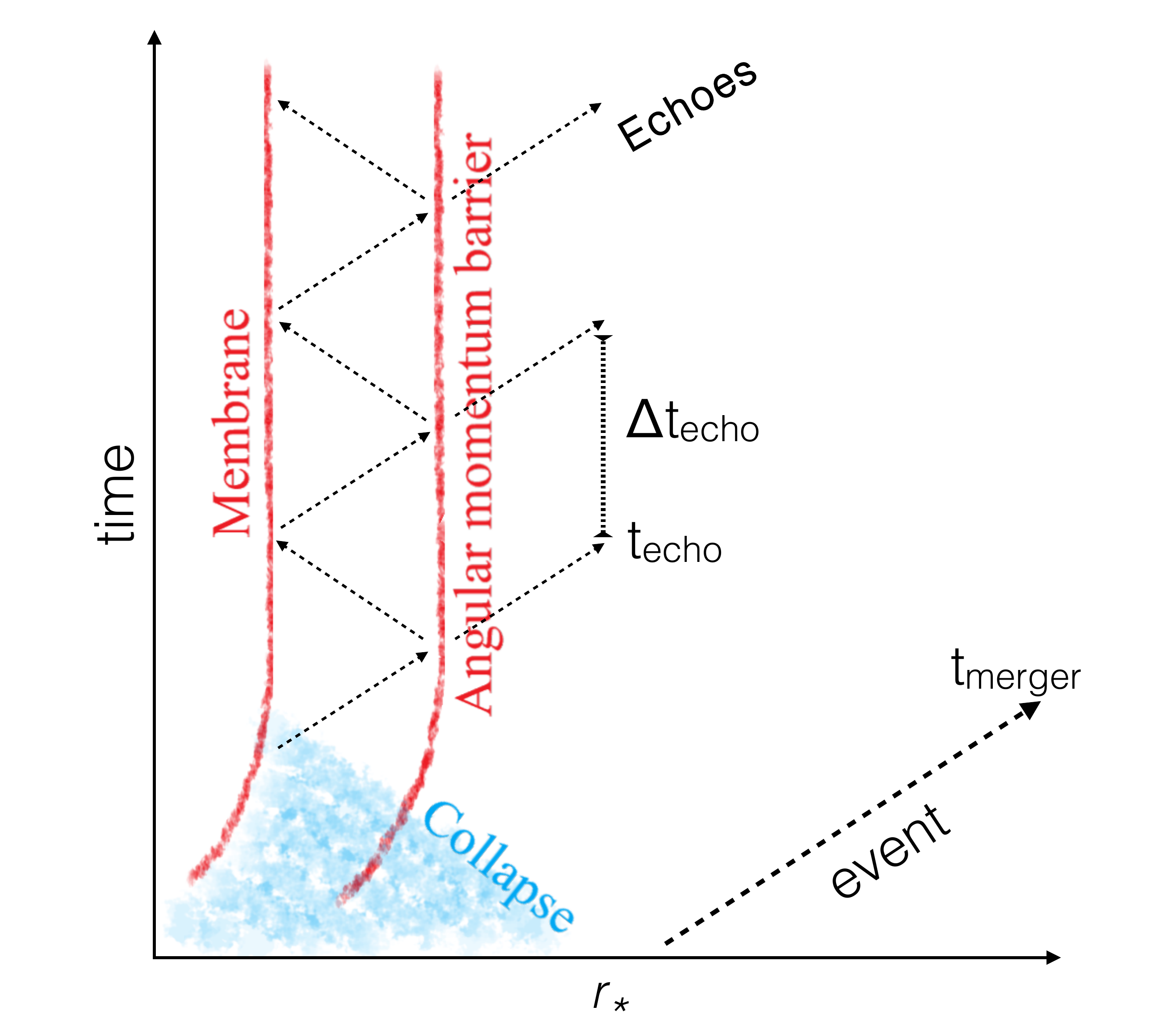}
\caption{Spacetime depiction of gravitational wave echoes from a membrane on the stretched horizon, following a collapse from binary neutron star merger event. }
\label{echo_pic_1}
\end{figure}

On the observational side, different groups have found tentative evidence (with p-values in the 0.2-2\% range) for repeating echoes in the LIGO/Virgo observations of binary black hole mergers \cite{Abedi:2016hgu,Conklin:2017lwb,Westerweck:2017hus}. While the instrumental vs. astrophysical origin of this signal remains controversial (see \cite{Ashton:2016xff, Westerweck:2017hus, Abedi:2017isz, Abedi:2018pst} for differing points of view), they can be interpreted as signatures of Planck-scale deviations near black hole horizons, which would validate the theoretical motivations based in the information paradox, or the cosmological constant problem. The echoes originate from the gravitational waves that are trapped between the angular momentum barrier and an exotic compact objects (ECOs) \cite{Cardoso:2016rao,Cardoso:2016oxy, Wang:2018gin} (see the next section for details), slowly leaking out and repeating within time intervals of:
\begin{eqnarray}
&\Delta t_{\rm echo} \simeq \frac{4 G M_{\rm BH}}{c^3}\left(1+\frac{1}{\sqrt{1-a^2}}\right) \times \ln\left(M_{\rm BH} \over M_{\rm planck}\right)& \nonumber\\ &\simeq 4.7~ {\rm msec} \left(M_{\rm BH} \over 2.7~ M_\odot \right) \left(1+\frac{1}{\sqrt{1-a^2}}\right),& \label{delay}
\end{eqnarray}
where $M_{\rm BH}$ and $a$ are the final mass and the dimensionless spin of the black hole remnant. Also, note that this time is similar to the scrambling time for quantum ``fast scramblers''  with entropy of the black hole  \cite{Hayden:2007cs, Sekino:2008he}.

Among the detected gravitational wave events, the first observation of a binary neutron star (BNS) coalescence GW170817 by the LIGO/Virgo gravitational wave detectors \cite{TheLIGOScientific:2017qsa,Abbott:2017dke} provides a unique opportunity to test general relativity, as well as the nature of the merger remnant, by searching for gravitational wave echoes.  
 After such a merger, a compact remnant forms whose nature mainly depends on the masses of the inspiralling objects. For GW170817, the final mass is within $\sim 2-3 M_{\odot}$ which could form either a black hole or a neutron star (NS). If it is a neutron star, it can be too massive for stability which means it will undergo a delayed collapse into a black hole. However, if a  black hole forms after the merger of GW170817, the ringdown frequency is outside the sensitivity band of LIGO/Virgo detectors, and thus they remain largely blind to this signal. Indeed, no post-merger GW signal in GW170817 has been detected \cite{Abbott:2017dke}. Nevertheless,  they can also lead to detectable echoes at lower freqeuncies in case an ECO is formed \cite{Cardoso:2016oxy}.

In nature, a BNS merger can develop in four possible ways: \cite{Abbott:2017dke}:
\begin{enumerate}
\item A black hole forms immediately after the merger.

\item A hypermassive NS is formed, then within $\lesssim$ 1 sec it collapses into a black hole.

\item A supramassive NS is formed that collapses to a black hole on timescales of $10$ - $10^{4}$ sec.

\item A stable neutron star is formed \cite{TheLIGOScientific:2017qsa}.

\end{enumerate}

A neutron star with mass greater than the maximum mass of a uniformly rotating star is hypermassive, as it is prevented from collapse through support from differential rotation and thermal gradients, caused by rapid cooling through neutrino emission.  
Eventually, magnetic braking of the differential rotation causes  such merger remnants to collapse within $\lesssim 1$  sec after formation \cite{Shapiro:2000zh,Hotokezaka:2013iia}.

\begin{figure}[!tbp]
\centering
    \includegraphics[width=0.7\textwidth]{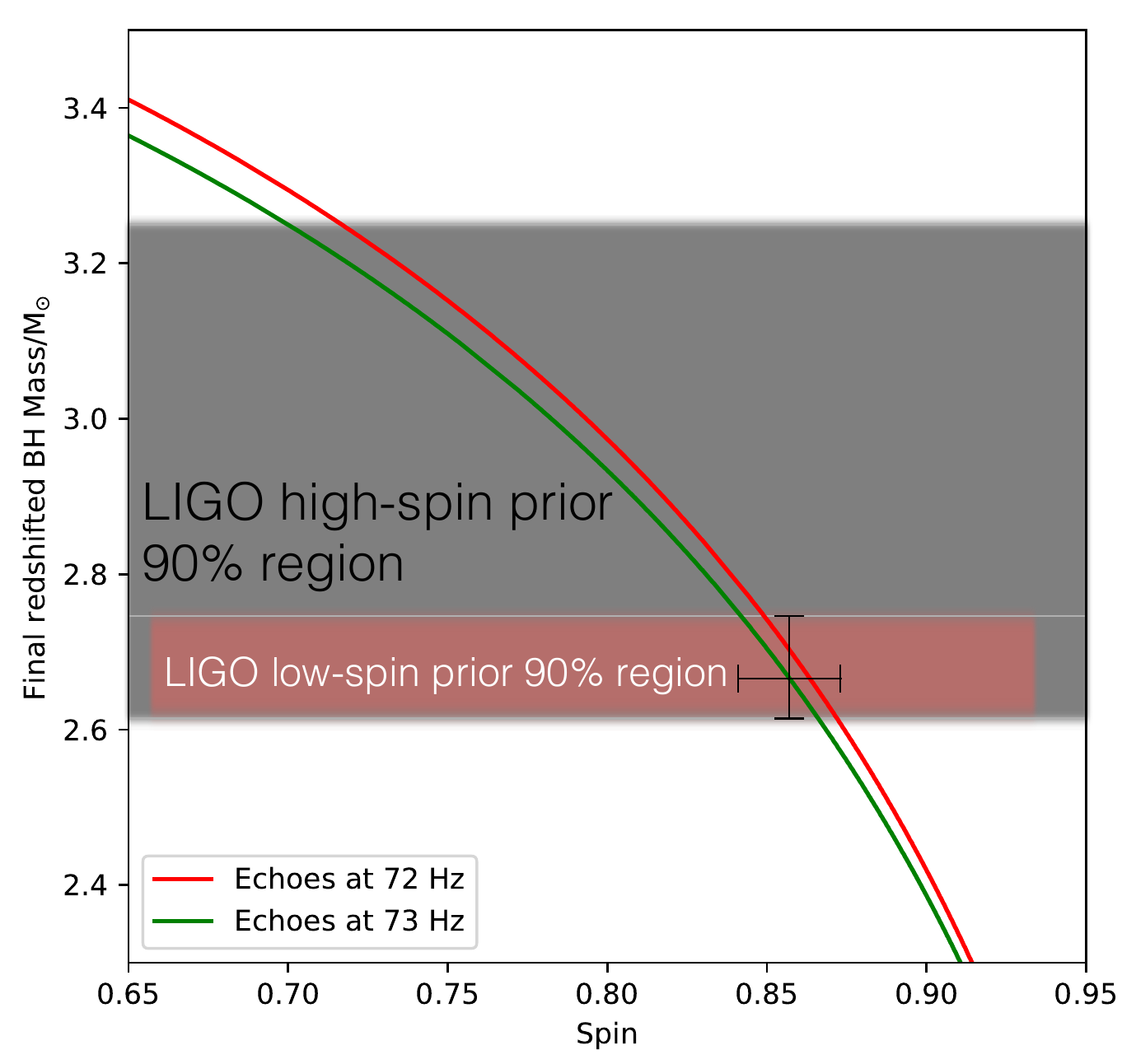}
 \caption{Constraints on mass and spin of the ``black hole'' remnant of GW170817 from the gravitational waves emitted during inspiral (shaded regions; assuming 2-5\% energy loss during merger) and post-merger echoes (lines). The low (high) BNS spin priors constrain the final dimensionless spin to be within 0.84-0.87 (0.70-0.87). }
 \label{NS-NS_12}
\end{figure}

For GW170817,  a broad range of NS equations of state lead to a post-merger mass which remains within the hypermassive NS regime \cite{TheLIGOScientific:2017qsa}. Accordingly, in this {\it paper} we present a search for short ($\lesssim$1 sec) duration GW echo signal potentially emitted from post-merger remnants in scenarios (1) or (2). We find tentative detection of echoes sourced by a ``black hole'' (or ECO) remnant of the neutron star collapse which is consistent with scenario (2) at $f_{\rm echo} \simeq 72$ Hz. 
Specifically, we find signatures of quantum gravitational alternatives to black hole horizons (as remnant of the BNS merger) in the gravitational wave data with statistical significance of $4.2\sigma$\footnote{In this {\it paper}, we use 1-tailed gaussian probability to assign a significance to a p-value, e.g., $1-$p-value$=$ 84\% and 98\% correspond to 1$\sigma$ and 2$\sigma$ respectively.}, or a false detection probability of $1.56 \times 10^{-5}$. This would constrain the spin of the ``black hole'' remnant to be within 0.84-0.87 (0.70-0.87) for the low (high) spin prior total BNS masses, reported by the LIGO/Virgo collaboration \cite{TheLIGOScientific:2017qsa} (see Fig. \ref{NS-NS_12}), consistent with the expected range from numerical simulations (e.g., \cite{Kastaun:2013mv}).

We shall first describe the expected properties of echoes, then our optimal search methodology, and finally results and discussions.

\section{Properties of echoes}\label{echo_prop}
Late time perturbations around Kerr black holes are often described in terms of  quasi-normal modes (QNM's). 
In classical GR, it is assumed that these modes are purely outgoing at infinity and purely ingoing  at the horizon. The continuous transition from ingoing to outgoing happens at the peak of the black hole angular momentum potential barrier. Correspondingly, as first noticed in \cite{Cardoso:2016rao,Cardoso:2016oxy}, if a membrane near the horizon (exotic compact object, or ECO) exists, it can partially reflect back the ingoing modes of the ringdown. In the ``geometric optics'' approximation, the reflected wavepacket reaches the angular momentum barrier after $\Delta t_{\rm echo}=8M\log M$ (+ spin corrections; see below) which is equal to twice the tortoise coordinate distance between the peak of the angular momentum  barrier ($r_{\rm max}$) and the membrane. The wavepacket would be partially transmitted through the barrier, appearing as delayed echo, and partially reflected, leading to subsequent echoes that repeat with a period of $\Delta t_{\rm echo}$ (see Fig. \ref{echo_pic_1}). 

These echoes have two natural frequencies: The resonance (or harmonic) frequencies of the ``echo chamber'' $f_{n} = n/\Delta t_{\rm echo}$ (also known as w-modes), where $n$ is an integer number, as well as the ``black hole'' ringdown (or QNM) frequencies that set the initial conditions for the echoes. The overlap of these frequencies will define the echo waveform. The higher harmonics ($n\gg 1$) that are primarily excited by the merger event decay quickly, while the lower harmonics are much more long lived (even marginally unstable \cite{Maggio:2017ivp,Bueno:2017hyj,Wang:2018gin}) as they are trapped by the angular momentum barrier. 

Since the final mass of BNS merger (2-3 $M_{\odot}$) is much smaller than that of the binary black hole mergers \cite{Abedi:2016hgu}, the lowest harmonic $ 1/\Delta t_{\rm echo}$ ($\simeq 80$ Hz), is shifted to the regime of LIGO sensitivity. This enables us to devise a model-agnostic strategy (such as in \cite{Conklin:2017lwb}) to search for lowest harmonics, independent of the initial conditions. 

\begin{figure*}[!tbp]
    \includegraphics[width=\textwidth]{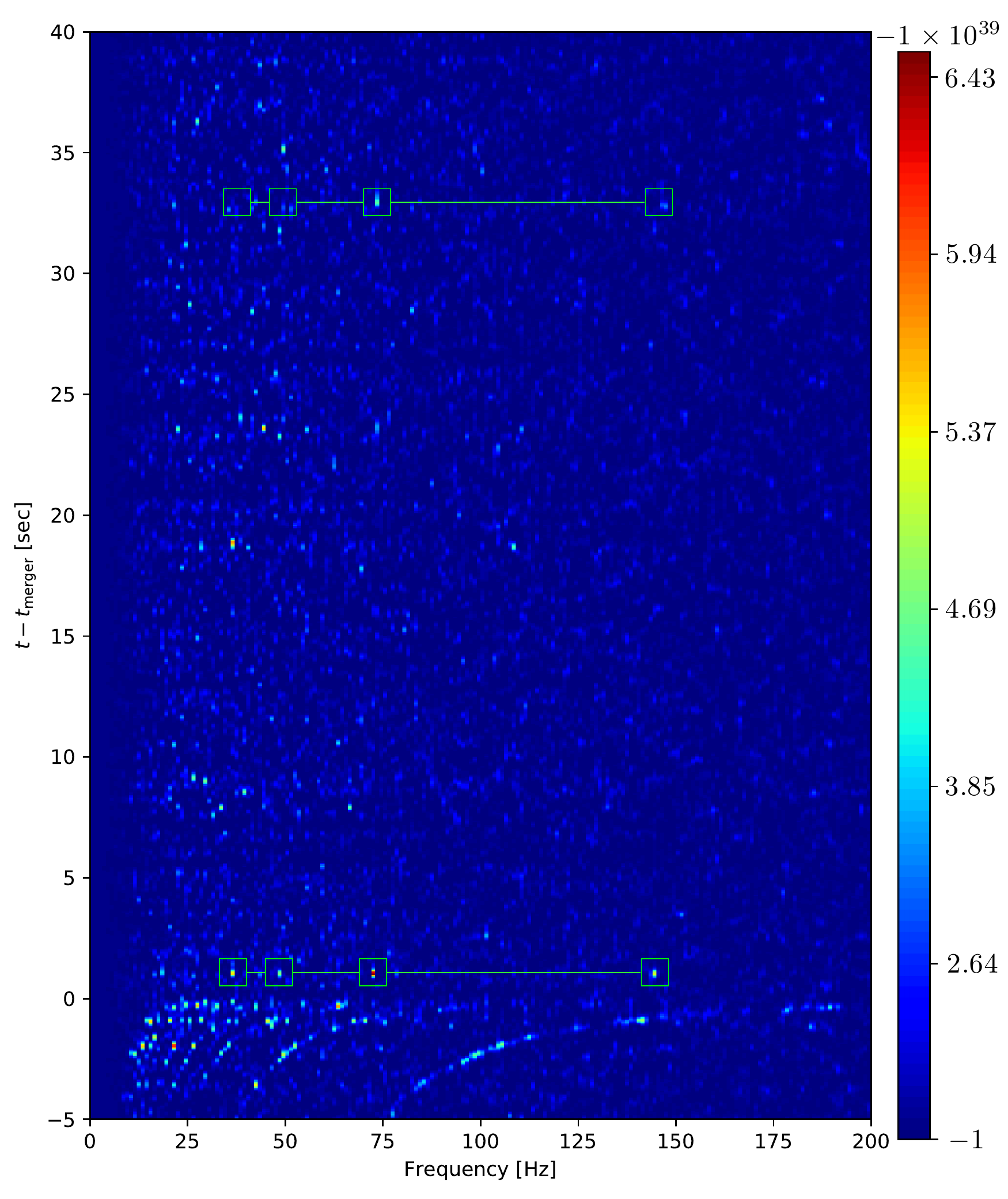}
 \caption{Time-frequency representations of $X(t,f)$ before and after the merger for the BNS merger gravitational-wave event GW170817, observed by the LIGO, and Virgo detectors. The possible peak of echoes found in the time-frequency and amplitude-frequency plots are marked with a green squares. As can be seen by the color scale, the peak at $f_{\rm{peak}}=72\ (\pm 0.5)$ Hz and $t-t_{\rm{merger}} \simeq $ 1.0 sec, with an amplitude of $X(t_{\rm{peak}},f_{\rm{peak}})=-6.48 \times 10^{39}$, is the highest peak in this diagram, from before {\it and} after the BNS merger (see Figs. \ref{NS-NS_3d}, \ref{NS-NS_4} and \ref{NS-NS_9} below for more details). A secondary tentative peak at the same frequency but $t-t_{\rm{merger}} \simeq $ 32.9 sec is also highlighted.}
 \label{NS-NS_11}
\end{figure*}

\begin{figure*}[!tbp]
    \includegraphics[width=\textwidth]{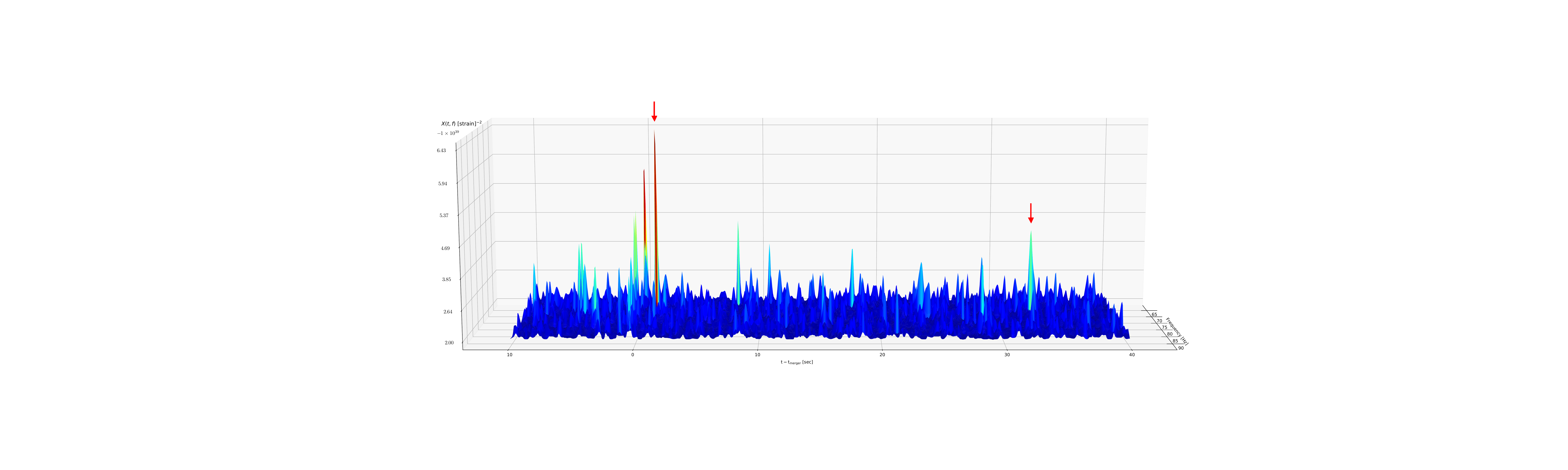}
 \caption{A 3d rendition of Fig. (\ref{NS-NS_11}) within our echo search frequency range $f=63-92$ Hz, showing that our tentative detection of echoes at  $f_{\rm{peak}}=72\ (\pm 0.5)$ Hz and $t-t_{\rm{merger}} \simeq $ 1.0 sec clearly stands above noise. }
 \label{NS-NS_3d}
\end{figure*}

\begin{figure*}[!tbp]
    \includegraphics[width=1\textwidth]{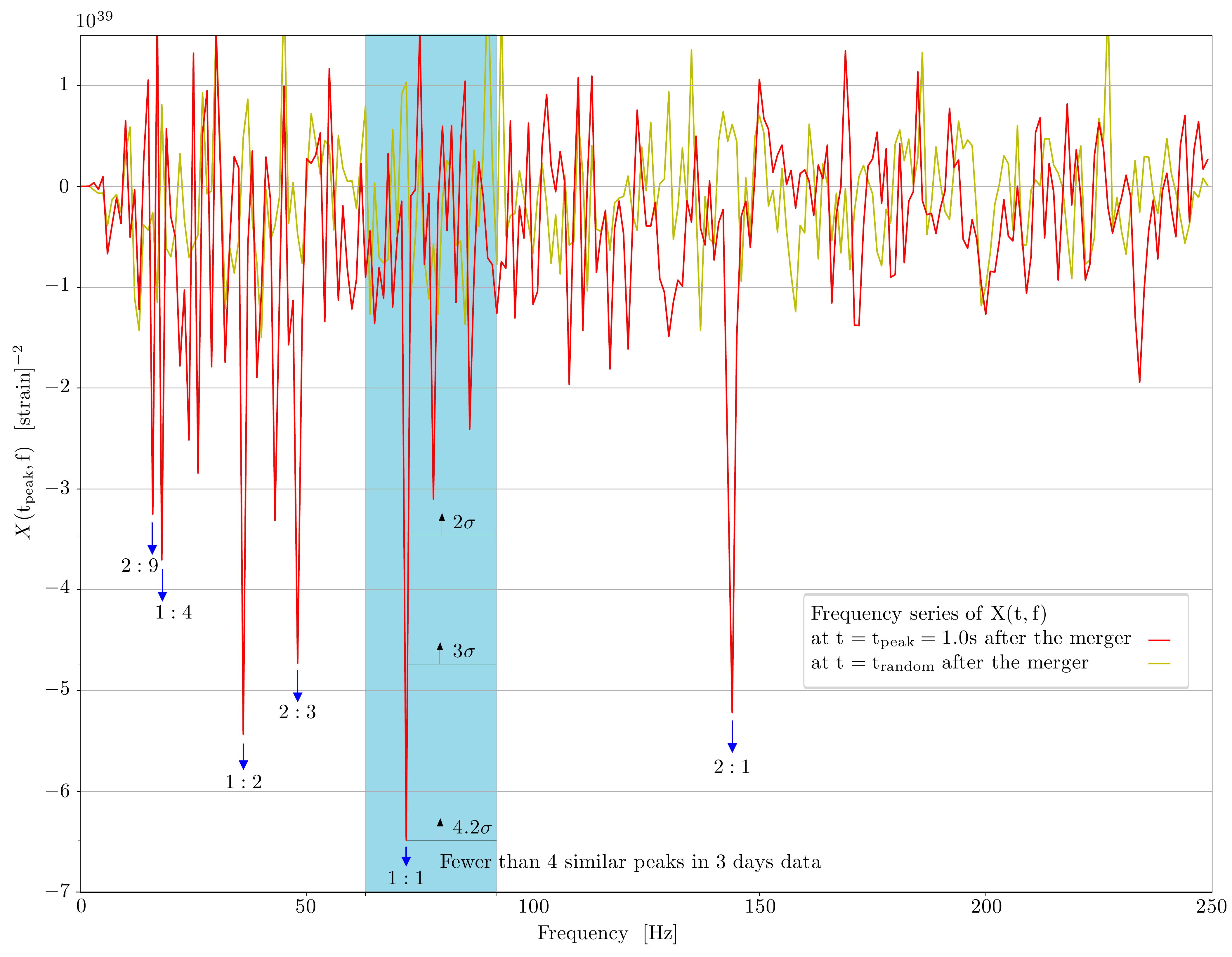}
 \caption{Amplitude-frequency representations of first peak (red) at 1.0 sec after the merger for the BNS merger gravitational-wave event GW170817, observed by x-correlating the LIGO detectors. Here the minimum of the first peak at 1.0 sec after the merger is $-6.48\times10^{39}$. We also plot the same amplitude-frequency for a random time in data (yellow). Other significant peaks associated with resonance frequencies are $\frac{2}{9}:\frac{1}{4}:\frac{1}{2}:\frac{2}{3}:1:2$ of 72 Hz.  Solid area between 63 Hz and 92 Hz was our prior range to search for Planckian echoes.}
 \label{NS-NS_4}
\end{figure*}

Geometric echo time delay for Kerr black hole with dimensionless spin parameter $a$ is given by \cite{Abedi:2016hgu}:
\begin{eqnarray}
\Delta t_{\rm echo}
=2r_{\rm max}-2r_{+}-2\Delta r 
+ 2\frac{r_{+}^{2}+a^{2}M^2}{r_{+}-r_{-}} \ln\left(\frac{r_{\rm max}-r_{+}}{\Delta r}\right) \nonumber \\
\ \ \ \ \ \ \ \ \ \ \ \ \ \ - 2\frac{r_{-}^{2}+a^{2}M^2}{r_{+}-r_{-}} \ln\left(\frac{r_{\rm max}-r_{-}}{r_{+}-r_{-}+\Delta r}\right),
\end{eqnarray}
where $r_{\pm}=M(1\pm \sqrt{1-a^{2}})$,  $\Delta r$ is the coordinate distance of the membrane and the (would-be) horizon, and $r_{\max}$ is the position of the peak of the angular momentum barrier   \cite{Abedi:2016hgu}. 

Placing the membrane at $\sim$ a Planck proper distance outside the horiozn: 
\be
\int_{r_{+}}^{r_{+}+\Delta r} \sqrt{g_{rr}} dr |_{\theta=0} \sim l_p \simeq 1.62 \times 10^{-33}~ {\rm cm},
\ee
 would be equivalent to setting $\Delta t_{\rm echo} \sim$ {\it scrambling time} for the black hole \cite{Hayden:2007cs,Sekino:2008he,Harlow:2014yka,Harlow:2013tf}. This choice can also be motivated by generic quantum gravity considerations such as the brick wall model \cite{THOOFT1985727},  trans-Planckian effects \cite{Khriplovich,York:1983zb}, 2-2 holes \cite{Holdom:2016nek}, or solving the cosmological constant problem \cite{PrescodWeinstein:2009mp}, which all fix the location of the membrane at $\Delta r|_{\theta=0} \sim{\sqrt{1-a^{2}} l_{p}^{2}}/{[4M(1+\sqrt{1-a^{2}})]}$ \cite{Abedi:2016hgu}. 


To find the prior for final BH mass, we use the  90\% credible region for the total (redshifted or detector frame) mass from Table I of \cite{TheLIGOScientific:2017qsa} (for the more realistic low-spin prior for BNS), but assume that $2-5\%$ of the energy is lost during the merger \cite{Zappa:2017xba}.  For the dimensionless BH spin $a=0.72-0.89$ (e.g., \cite{Kastaun:2013mv}), as well as an order of magnitude change in Planck length, we can constrain $\Delta t_{\rm echo}$ to be within:
\begin{eqnarray}
0.0109<\Delta t_{{\rm echo}}({\rm sec})<0.0159.
 \end{eqnarray}
This range in echo time delay corresponds to echo frequency ($f_{\rm echo} = \Delta t_{\rm echo}^{-1}$) range:
\begin{eqnarray}
63 \leq f_{{\rm echo} }({\rm Hz}) \leq 92, \label{f_prior}
 \end{eqnarray}
which we shall use for our echo search and background estimates, for the rest of the paper.

\section{Model-agnostic search for echoes}\label{method}

The simplicity of a method indicates its degree of certainty and minimizes any ambiguity. Motivated by the model-agnostic echo search proposed in \cite{Conklin:2017lwb}, here we devise an optimal (minimum variance) method to extract periodic gravitational wave bursts in the cross-correlation of the two detectors:
\be
h(t) \propto \sum_n \delta_D(t-n\Delta t_{\rm echo} -t_0),
\ee 
or
\be
h_{f} \propto \sum_n \delta_D(f- n f_{\rm echo}), \label{h_f_res}
\ee
in frequency space, $\delta_D$ is the Dirac delta function. This is clearly a simplification, but can be justified as:
\begin{enumerate}
\item The structure of individual echoes cannot be resolved, as they are set by the QNM frequency that is beyond the LIGO sensitivity window for the BNS masses.
\item Echoes decay slowly in the low frequency and/or superradiance regime \cite{Wang:2018gin}.
\end{enumerate}

Therefore, given that the noise is uncorrelated between the two detectors while the signal should be the same, we search for periodic peaks of equal amplitude in the cross-power spectrum of the two detectors at integer multiples of $f_{\rm echo}$.

In this analysis, we use the observed strain datastream of GW170817 \cite{GW170817} occured at GPS time $t_{\rm{merger}}=$ 1187008882.43 == August 17 2017, 12:41:04.43 UTC from the two detectors Hanford and Livingston. We call these $h_{H}(t)$ and $h_{L}(t)$, respectively. We used the strain data at 16 kHz and for 2048 sec duration. First we calculate the offset between detectors. By correlating the two datasets (before the merger) we obtain that for GW170817, Hanford event is $\delta t \simeq 2.62$ msec delayed with respect to Livingston (see \ref{delta_t} for details). This is consistent with what has already been reported \cite{GBM:2017lvd}. Similarly, we can confirm the relative polarization of $-1$  between the two detectors. 

Based on general properties of echo signal outlined above, we build our model-agnostic search as follows:
\begin{enumerate}
\item We vary the fundamental frequency of echoes  $f_{\rm echo}= \Delta t_{\rm echo}^{-1}$ within the 90\% credible region range, as outlined in Sec. \ref{echo_prop}:
$$
63 \leq f_{{\rm echo} }({\rm Hz}) \leq 92.
$$
We shall then refer to $f_{\rm{peak}}$ as the best fit value of $f_{\rm echo}$, which maximizes ($-1\times$) cross-correlation of the two detectors. 



\item We search for the echo signal within the range $0 <t-t_{\rm{merger}} \leq 1~ \rm{sec}$. This is the prior range already used by Abbott et al. 2017 \cite{Abbott:2017dke} in search of post-merger gravitational waves from remnant of BNS merger GW170817.

\item We obtain the amplitude spectral density (ASD) with the same method introduced in \cite{GW150914} which was used for whitening the data. As we are interested to combine different frequencies to obtain the echo signal, we Wiener filter (rather than whiten) the data by dividing by noise variance PSD=ASD$^2$ (rather than ASD). This ensures the optimal combination of different frequencies, which receives {\it minimum variance} from noise:
\begin{eqnarray}
&H(t,f)=\rm{Spectrogram}\left[ \rm{IFFT}\left( \frac{\rm{FFT}(h_{H}(t-\delta t))}{PSD_{H}} \right) \right],& \nn\\ 
&L(t,f)=\rm{Spectrogram}\left[ \rm{IFFT}\left( \frac{\rm{FFT}(h_{L}(t))}{PSD_{L}} \right) \right].& \nn\\ &&
\end{eqnarray}
Here the same setup as LIGO code \cite{GW150914} to obtain the spectrogrm with mlab.specgram() function in Python with NFFT=$f_{s}=16384$, and mode='complex' is used.  Same as the method used in \cite{GW150914}, the number of points of overlap between blocks is NOVL = $NFFT \times 15/16$. 
 With weighting the frequencies and after a right normalization the resulting time series becomes in units of strain$^{-1}$. For consistency checks, instead of mlab.specgram() we also used one-dimensional discrete Fourier transform for real input np.fft.rfft() taking 1 sec data segments, with steps of 1/16 of a second, and after right normalization we reproduced same results. As another consistency check, same results is reproduced using 4096 Hz data. Here the ASDs are obtained for all the 2048 sec range of data.

\item We then cross-correlate the resulting spectrograms  and sum over all the resonance frequencies of $nf$. More specifically, we define 
\begin{eqnarray}
X(t,f)=\sum_{n=1}^{10} \Re\left[H(t,nf) \times L^*(t,nf)\right], \label{x_def}
\end{eqnarray}
which is the Wiener-filtered cross-power spectrum of the two detectors for the echo signal \ref{h_f_res}. The choice of $n\leq 10$ does not affect our results, as the LIGO noise (or ASD$^2$) blows up at high frequencies, and thus they do not contribute to $X(t,f)$. Due to the opposite polarizations of the LIGO detectors mentioned above, the real gravitational wave signals show up as peaks in $-X(t,f)$.

\item For obtaining the p-value we used 2048 sec data of GW170817 at 16384 Hz after noise subtraction. We picked 1500 sec = 25 min of this data ($\sim (-1600,-100)$ sec before time of merger). Then with shifting detectors we produced $\sim$14 days of ``data'' out of the original 25 minutes. This data is ``time-shifted'', i.e. a non-physical time lag is introduced between the detectors analyzed so as to remove correlated gravitational-wave signals. This data is also ``off-source'', i.e. it is outside of the 100 sec window of event. Assuming that the detector noise properties remain reasonably steady throughout the observation time, this analysis produces independent realizations of noise (or background) for $X(t,f)$, which can be used to estimate the false-alarm probability for a given possible detection.

\end{enumerate}

The strength of this method is in the simplicity of the computation, which only involves a straightforward x-correlation in frequency space,  without any arbitrary or {\it ad-hoc} cuts or parameters.

\section{\label{Results}Results}

\begin{figure}[!tbp]
\centering
    \includegraphics[width=0.8\textwidth]{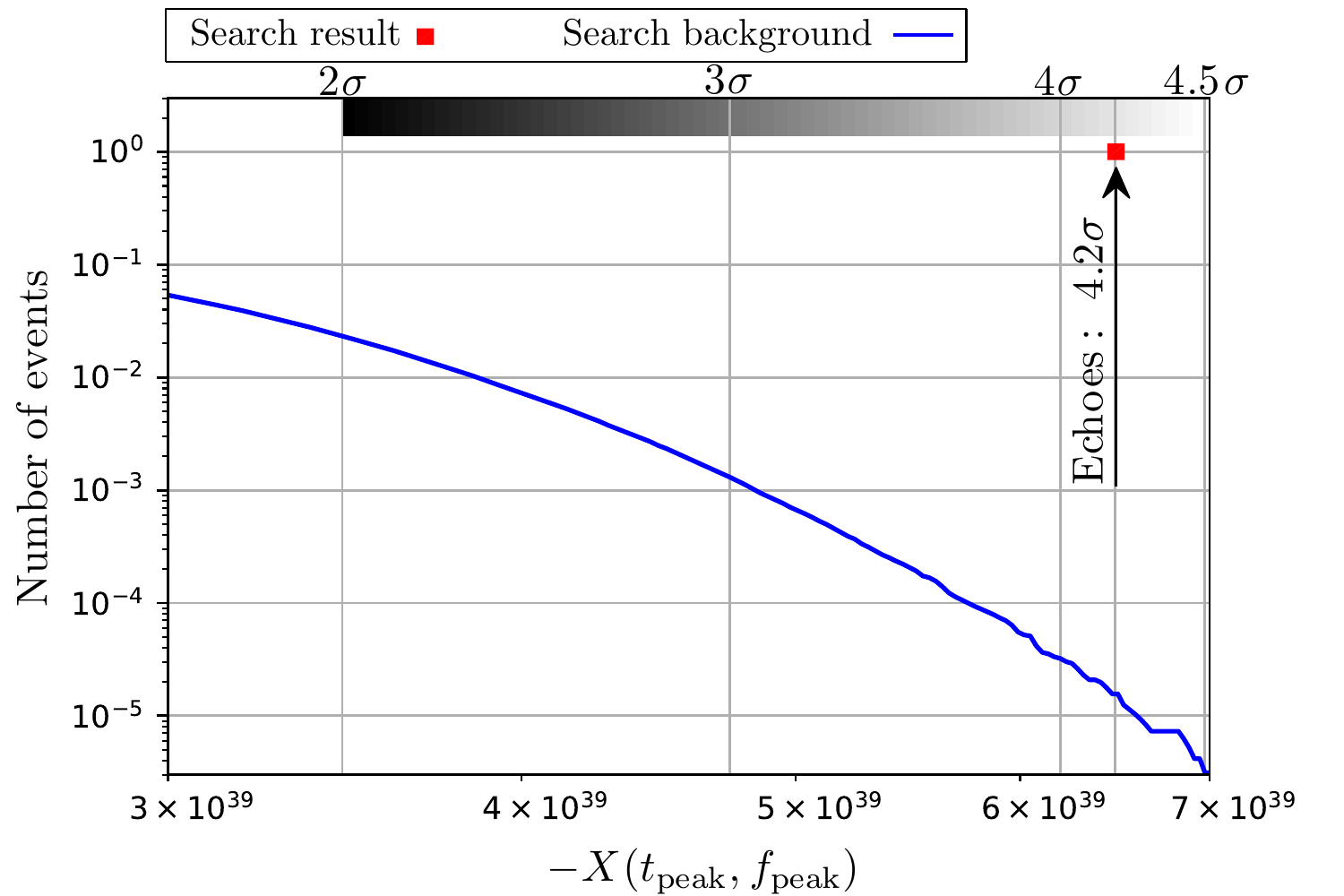}
 \caption{Average number of noise peaks higher than a particular -X(t,f) within a frequency-intervals of 63-92 Hz and time-intervals of 1 sec for LIGO noise near GW170817 event. The red square shows the observed $-X(t_{\rm{peak}},f_{\rm{peak}})$ peak at 1.0 sec after the merger. The horizontal bar shows the correspondence between $X(t,f)$ values and their significance. This histogram obtained from producing $\sim$2 weeks data out of off-source 2048 sec available data \cite{GW170817}.}
 \label{NS-NS_10}
\end{figure}

\begin{figure}[!tbp]
\centering
    \includegraphics[width=0.7\textwidth]{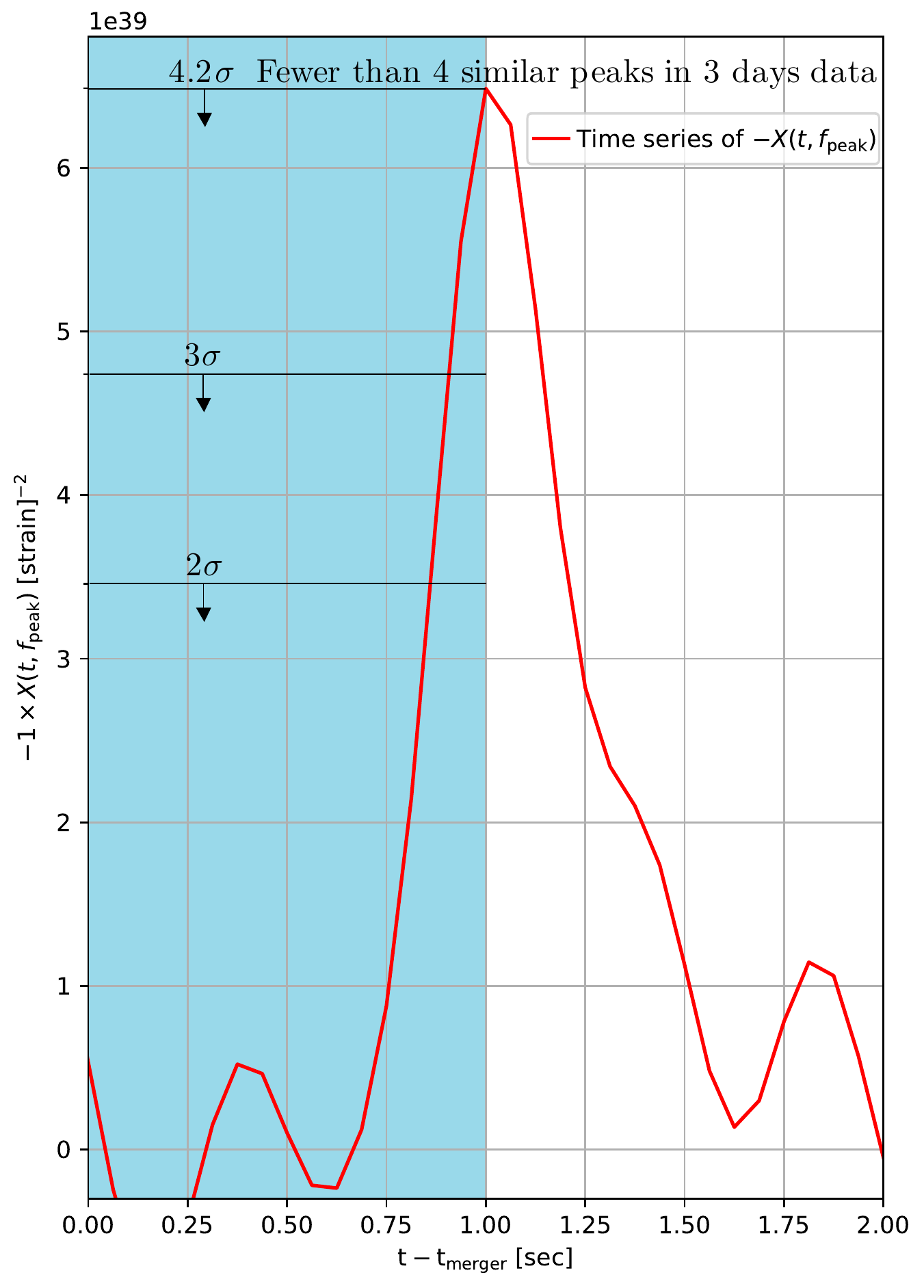}
 \caption{Amplitude-time representations of first (and most significant) echo peak at 1.0 sec after the merger and frequency of 72 Hz. The shaded region is 0-1 sec prior range after the merger, first adopted in \cite{Abbott:2017dke}, which we use to estimate p-value . The maximum of the peak is $6.48\times10^{39}$.}
 \label{NS-NS_9}
\end{figure}

We now present the outcome of the search for gravitational wave echo signal from the remnant of the BNS merger GW170817 using data from LIGO/Virgo collaboration, within  $\lesssim 1$ sec after the merger. The signal could be a consequence of a hypermassive NS collapse, which has collapsed to form a ``black hole'' (or ECO) \cite{Abbott:2017dke}.
We find a significant negative peak in the spectrogram in Figs. \ref{NS-NS_11}, \ref{NS-NS_3d} and \ref{NS-NS_4}, at $t-t_{\rm{merger}} \simeq $ 1.0 sec after the merger (see Fig. \ref{NS-NS_9}), at $f_{\rm echo} \simeq 72$ Hz. As it is shown in Fig. \ref{NS-NS_4}, it is also accompanied by secondary resonances, as all the periodic signals (Eq. \ref{h_f_res}) with a rational ratio of periods, have a non-vanishing overlap (also see \cite{Conklin:2017lwb} for a similar effect). Accounting for the ``look elsewhere'' effect, we find tentative detection of Planck-scale structure near black hole horizons at false detection probability of $1.56\times 10^{-5}$ (corresponding to $4.2\sigma$ significance level in Fig. \ref{NS-NS_10}). This means fewer than 4 similar peaks can be found in three days of ``data'', as discussed in the Sec. \ref{method}. 

As we hinted in the Introduction above, the frequency of this peak would constrain the dimensionless spin of the ``black hole'' or ECO remnant to be within 0.84-0.87 (0.70-0.87) for the low (high) spin prior total BNS masses \cite{TheLIGOScientific:2017qsa} (see Fig. \ref{NS-NS_12}). For the more realistic low-spin prior on BNS spins, the tight constraint of $a=0.84-0.87$  is exactly in line with numerical simulations of BNS mergers (e.g., \cite{Kastaun:2013mv}), as opposed to $a\simeq 0.67$ for binary BH mergers (e.g., \cite{TheLIGOScientific:2016pea}). 

\begin{figure*}[!tbp]
    \includegraphics[width=1\textwidth]{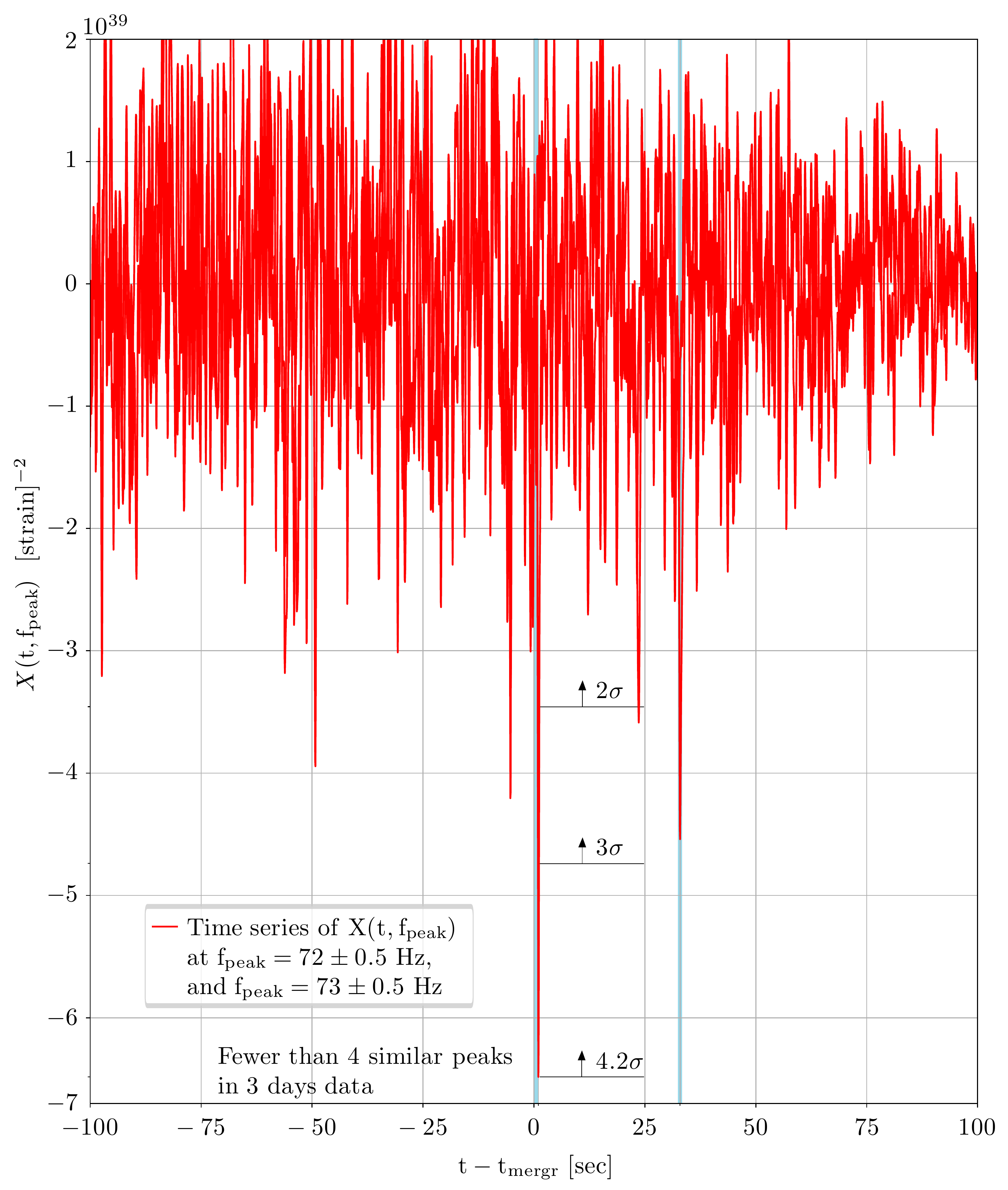}
 \caption{Amplitude-time representations of peaks before and after the merger $t-t_{merger}=$(-100s,100s) for the BNS merger gravitational-wave event GW170817, observed by the LIGO, and Virgo detectors. As highlighted, the first and second highest negative peaks appear 1.0 sec and 32.9 sec after the merger respectively.  This plot is for fixed frequencies of 72 ($\pm0.5$) Hz and 73 ($\pm0.5$) Hz. First solid area is 1 sec prior range after the merger determined in \cite{Abbott:2017dke}. As can be seen minimum of first peak is $-6.48\times10^{39}$ and minimum of second peak is $-4.54 \times 10^{39}$.}
 \label{NS-NS_2}
\end{figure*}

\begin{figure*}[!tbp]
    \includegraphics[width=1\textwidth]{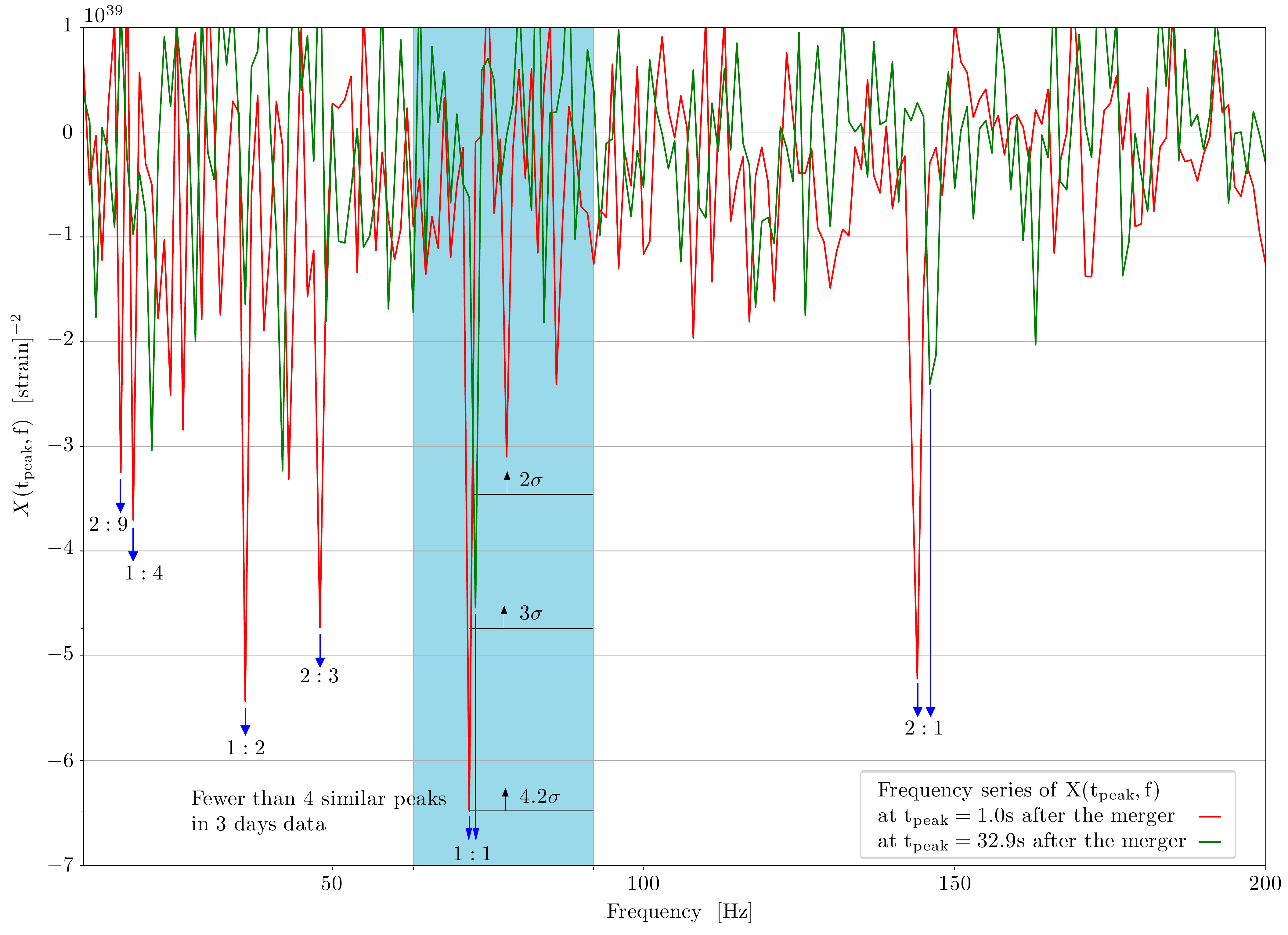}
 \caption{Amplitude-frequency representations of data after the merger for the peaks at 1.0 sec (red) and 32.9 sec (green) after the merger for the BNS merger gravitational-wave event GW170817, observed by the LIGO, and Virgo detectors. Here the resonance frequencies for the first peak at 1.0 sec and second peak at 32.9 sec after the merger are shown in this plot. For the second peak there is identical pattern with a factor 1.0139 in these frequencies. Here the maximum of first peak is $-6.48\times10^{39}$ and maximum of second peak is $-4.54 \times 10^{39}$. Solid area between 63 Hz and 92 Hz is the region we expect to see Planckian echoes.}
 \label{NS-NS_3}
\end{figure*}

Along with this significant peak, there exists a lower peak with nearly identical resonance pattern at 73 Hz and $t-t_{\rm{merger}}$ = 32.9 sec after the merger (see Figs. \ref{NS-NS_11}, \ref{NS-NS_3d}, \ref{NS-NS_2} and \ref{NS-NS_3}). While this secondary peak could be due to detector noise, it might also originate from post-merger infall into the ``black hole'' or ECO, which would cause a slight change in spin and mass.

The observed peak at $t-t_{\rm{merger}} \simeq$ 1.0 sec is consistent with Scenario 2 in Section \ref{Introduction}, which is a hypermassive NS remnant \cite{Abbott:2017dke} which collapses into a BH within $\sim$ a second. Interestingly, this merger might lead to the formation of a torus around this black hole, providing the sufficient power for a short $\gamma$-ray burst (GRB) \cite{Kastaun:2013mv} from its rotational energy, fuelling a jet that has been detected (170817A) by the Fermi Gamma-ray Burst Monitor \cite{Goldstein:2017mmi,Monitor:2017mdv} 1.7 sec after the merger.

\section{\label{discuss} Discussions} 

In this section, we shall go over some sanity checks: 
\begin{enumerate}

\item The most immediate question on reader's mind might be whether our tentative detection of echoes at $72$ Hz, around 1.0 second after the merger, might be due to some type of leakage from BNS inspiral signal. Now, this signal is computed from x-correlation of detectors within $0.5 ~{\rm sec}<  t - t_{\rm merger} < 1.5 ~{\rm sec}$, and thus is not expected to be affected by the pre-merger inspiral signal, or the LIGO-Livingston glitch at 1.1 second before merger. Furthermore, the most significant peak in $X(t,f)$ in Figs. (\ref{NS-NS_11}) and (\ref{NS-NS_3d}) is indeed our echo peak at 72 Hz and 1 sec, while the pre-merger signal is less significant (at any particular frequency), and does not appear to have any specific feature at 72 Hz. Furthermore, recurrence of the same frequency at 33 seconds after the merger makes the possibility of leakage from before merger less likely.

\item Fig. \ref{whitened} shows the simple cross-power spectrum of whitened, time-shifted data from the two detectors, around the time of observed echo peaks, 1.0 sec and 32.9 sec after the merger. Given that, within the assumption of gaussian detector noise, the noise should be uncorrelated and uniform on this plot, we can assess the significance of various harmonics. The vertical bands show the integer multiples of our tentative detection $f_{\rm echo}({\rm Hz}) = 72.0 \pm 0.5$. We see that different harmonics have different levels of excitement at different times. Interestingly, the 2nd (1st) harmonic has the most significant amplitude for 1st (2nd) echo peak. Moreover, for the 1st (more significant) echo peak at 1.0 sec, we switch from even harmonics 2 and 4, to odd ones, i.e. 7 and (possibly) 9. There is also a curious peak at 48 Hz = $\frac{2}{3} f_{\rm echo}$ (for both echo peaks), which is {\it not} captured by the linear echo model (\ref{h_f_res}) and may indicate resonance with nonlinear interactions, e.g., $h^3$ at the source. 

\item Another worry is the leakage from LIGO calibration lines at 34 Hz-37 Hz, which are nearly 1/2 of our observed $f_{\rm echo} = 72$ Hz\footnote{We thank Ofek Birnholtz for bringing this point to our attention}. While this is a possibility, it is not clear why this would only happen at 1 second after BNS merger, and not two weeks of pre-merger ``data'' we use for our background search. In fact, our most significant peaks at 144 Hz and 288 Hz in the whitened cross-power spectrum  (see Fig. \ref{whitened}) do not appear to correspond to any particular feature in the LIGO noise curve. Moreover, the most significant x-correlation peaks in Fig. \ref{whitened} are all negative, as expected for real gravitational wave signal from GW170817, while there is no physical process that would correlate the detector noise from Hanford and Livingston detectors.  

\item Phase shifts upon reflections off the angular momentum barrier, or the ECO/membrane may shift the natural frequencies of the echo\footnote{We thank Jing Ren for bringing this point to our attention.}. At frequencies much lower than the QNM frequencies ($\sim 10^4$ Hz; associated with 1/thickness of the angular momentum barrier, or the Hawking temperature associated with the ``firewall'' or ``fuzzball''), we expect the phase shifts, and thus the shift in natural frequencies to approach a constant, i.e. 
\begin{equation}
f_{n} = \left(n- \frac{\phi_1+\phi_2}{2\pi}\right)f_{\rm echo} \simeq n f_{\rm echo} + {\rm const.},   
\end{equation}
where $\phi_1$ and $\phi_2$ are phase shifts at the angular momentum barrier and the ECO/membrane, respectively. However, note that this does not change the spacing of the $f_n$'s. The spacing of our two most significant frequency peaks in Fig. \ref{whitened} (1 second after the merger) is $\Delta f \simeq 288 ~{\rm Hz} - 144~ {\rm Hz} = 144$ Hz. Now, considering the frequencies within our prior range $f_{\rm echo} =63-92$ Hz (Eq. \ref{f_prior}), the only frequency that can be multiplied by an integer to give $\Delta f$, is $f_{\rm echo} = 72$ Hz, consistent with our prior finding. This also suggests that the sum of the phase shifts $\phi_1+\phi_2$ must be an integer multiple of $2\pi$ (at least, within a couple of percent).  

\item Let us next estimate the gravitational wave energy emitted in echoes. 
%
%
%
%
%
%
We start by noticing that the negative peaks at $t-t_{\rm merger} \simeq -0.9$ and $-5.4$ sec prior to the merger in Fig. \ref{NS-NS_2}, correspond to the inspiral signal at 144 and 72 Hz, respectively (see also Fig. \ref{NS-NS_11}-\ref{NS-NS_3d}). We now can use the energy emitted during the inspiral:
\be
\Delta E_{GW} = -\frac{{\cal M}}{2}\Delta (\pi f G {\cal M})^{2/3}, 
\ee  
in terms of chirp mass ${\cal M} \simeq 1.188 ~M_\odot$ \cite{TheLIGOScientific:2017qsa}, to estimate that the fraction of inspiral energy radiated within 72-73 Hz and 144-145 Hz, compared to all energy emitted in LIGO band ($\sim 0.025~ M_\odot c^2$) is 0.5\%. We would gain an extra factor of $\sim 3$, by including the 4th harmonic at 288 Hz, which is the only other significant harmonic present in data up to 500 Hz  (given that GW energy scales at $f^2 \times$ amplitude$^2$).  Since the amplitudes of inspiral peaks in Fig. \ref{NS-NS_2} are a factor of $\sim 2$ smaller than the echoes (within 1 Hz frequency windows), we can estimate that the fraction of radiated post-merger echo to radiated BNS inspiral energy (within LIGO sensitivity band) is:
\be
\frac{E_{\rm echo}(\lesssim 500 ~{\rm Hz})}{E_{\rm inspiral}(\lesssim 500 ~{\rm Hz})} \sim 3\%,
\ee
for the first and most prominent echo peak $\lesssim 1$ sec after the merger. Any estimate beyond this frequency range will be highly model-dependent. 

\begin{figure*}[!tbp]
    \includegraphics[scale=0.5,center]{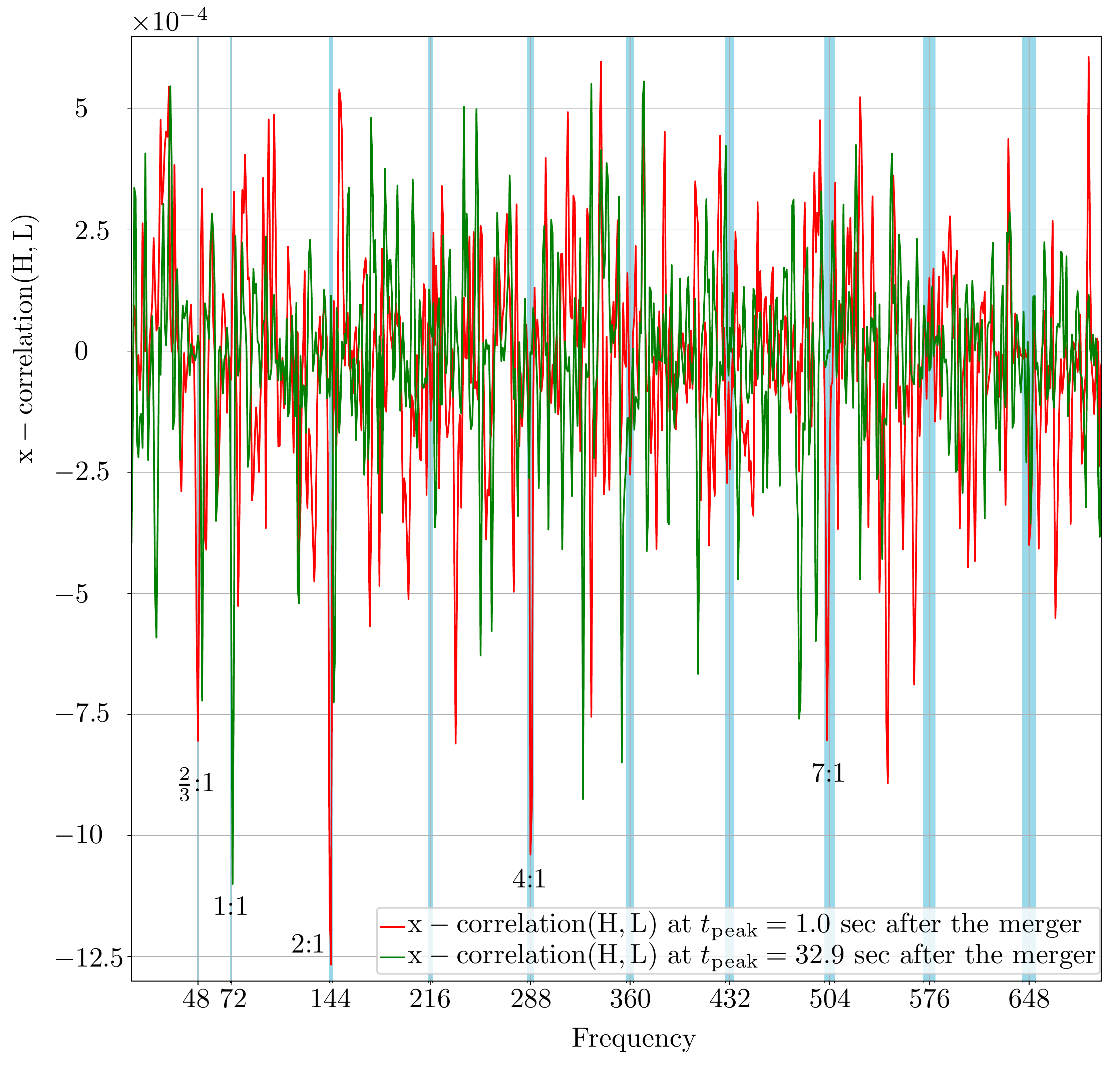}
 \caption{The simple cross-power spectrum of Hanford and Livingston detectors (with a 2.62 msec time delay), after whitening each dataset independently at $t-t_{\rm merger}({\rm sec}) = 1.0$ (red) and $32.9$ (green). The vertical bands show all the integer multiples of $f_{\rm echo}({\rm Hz}) = 72.0 \pm 0.5$ in frequency (which contribute to $X(t,f_{\rm echo})$ in Eq. \ref{x_def}), as well as a curious peak at $\frac{2}{3} f_{\rm echo}$. Different harmonics of the ``echo chamber'' appear to be excited at different times  }
 \label{whitened}
\end{figure*}

\begin{figure*}[!tbp]
    \includegraphics[width=1\textwidth]{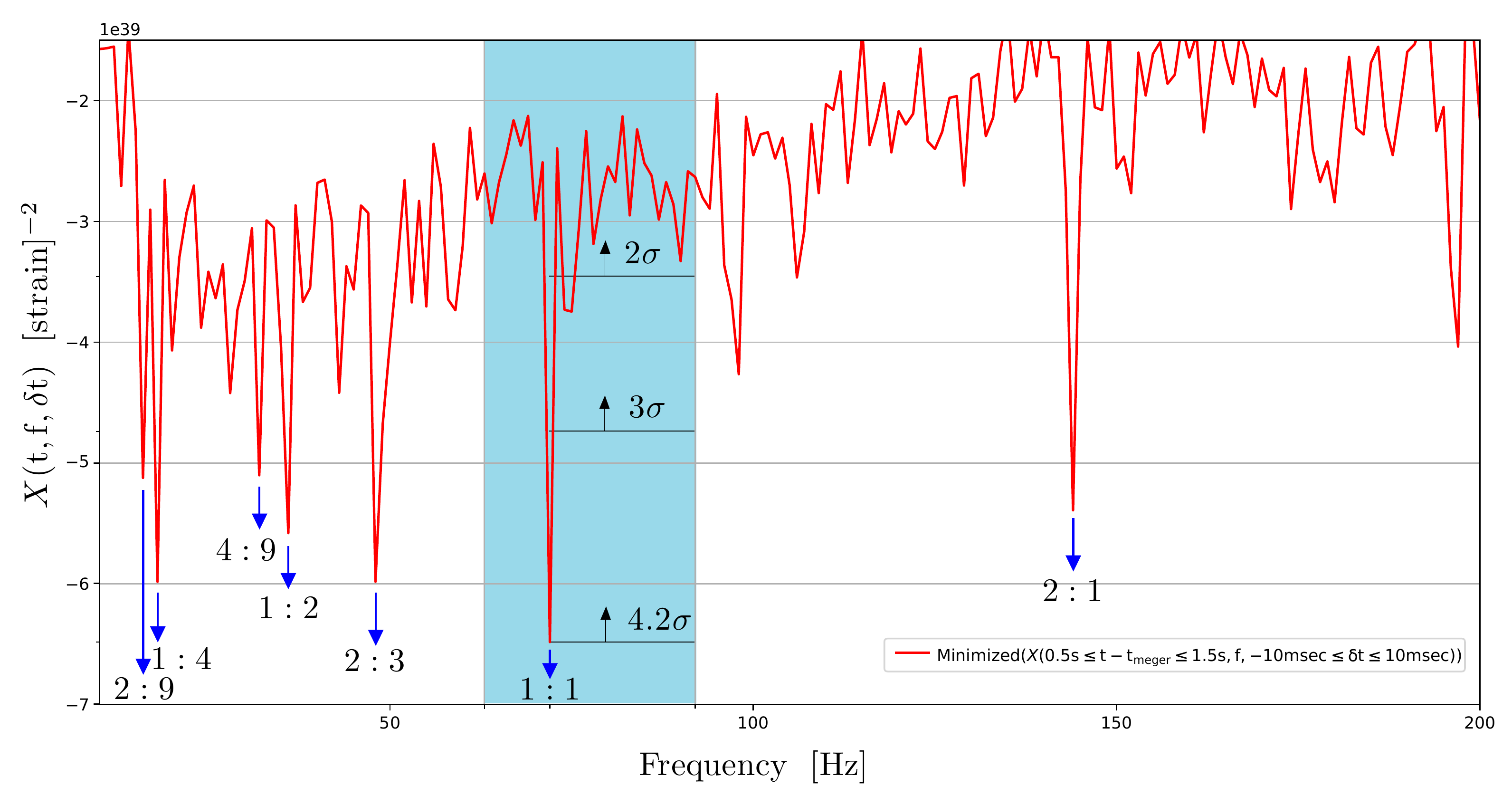}
 \caption{Amplitude-frequency representations peak of $\rm{t_{peak}}$=1.0 sec after the merger. In this analysis the detectors are shifted within $\pm 10$ms and the $X(t,f)$ is minimized over ($\rm{t_{peak}}$-0.5,$\rm{t_{peak}}$+0.5) sec data range after the merger. Here the minimum peak again is the one with 72 Hz frequency, which happens at 2.62 ms time  shift between Hanford and Livingston detectors, consistent with main event (see \ref{delta_t}). Resonance frequencies are at $\frac{2}{9}:\frac{1}{4}:\frac{4}{9}:\frac{1}{2}:\frac{2}{3}:1:2$ of 72 Hz.}
 \label{NS-NS_7}
\end{figure*}

\begin{figure*}[!tbp]
    \includegraphics[width=1\textwidth]{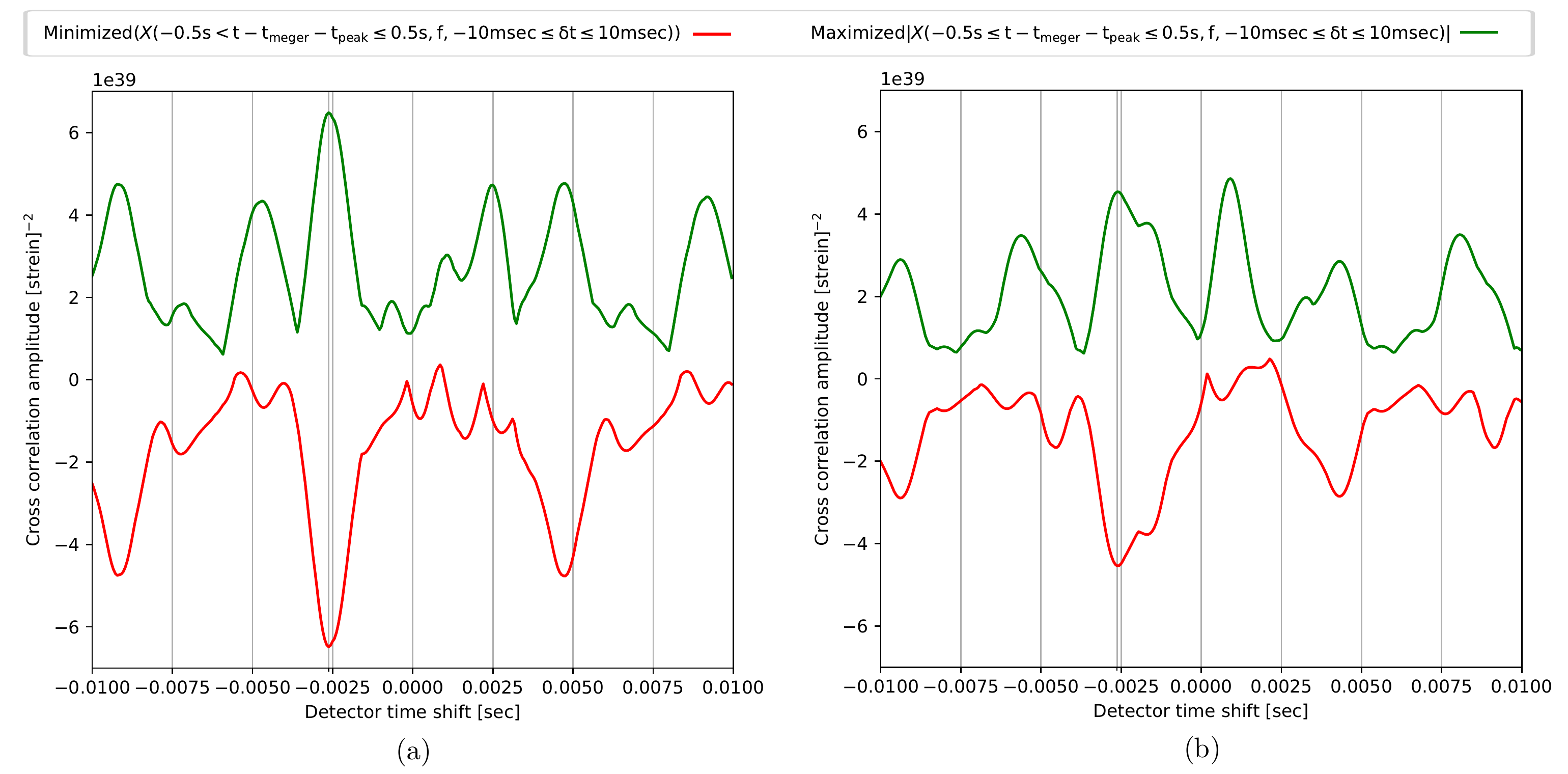}
 \caption{(a) Peak of 1.0 sec after the merger. (b) Peak of 32.9 sec after the merger. We see that for both of the peaks the minimums (red), maximums (green) are at 2.62 ms. Minimized (red) Maximized (green) $X(t,f)$ representations of data after the merger within $-0.5s<t-t_{\rm{merger}}-1.0s<0.5s$ and $-0.5s<t-t_{\rm{merger}}-32.9s<0.5s$ for varying shift parameter 10 ms$<\delta t<$10 ms for the BNS merger gravitational-wave data GW170817. Since we have resonance frequencies we see a periodic pattern.}
 \label{NS-NS_5}
\end{figure*}
%

\item In order to further test the consistency of the signal, we shifted detectors within the range of (-10 ms, 10 ms) and minimized $X(t-t_{\rm{merger}},f)$ within the range $0.5 ~{\rm sec}<t-t_{\rm{merger}}\leq 1.5 ~{\rm sec}$. Fig. \ref{NS-NS_7} shows this minimization and confirms that the negative peak at $f_{\rm{peak}}$=72 Hz is indeed the minimum. Similarly, Figs. \ref{NS-NS_5} (a)-(b)  confirm that the peaks at $-0.5<t_{\rm{peak}}({\rm sec})-1.0<0.5$, and $-0.5<t_{\rm{peak}}({\rm sec})-32.9<0.5$ show a minimum when detector times are shifted by 2.62 ms, which is exactly the same time delay seen for the main event \cite{TheLIGOScientific:2017qsa}. Moreover, the plots confirm that relative polarization of the two detectors which are opposite to each other for the main event, as well as the peaks at 1.0 sec and 32.9 sec after the merger. 


\item During the course of this work, an updated version of \cite{Conklin:2017lwb} appeared on arXiv, claiming evidence for an echo frequency of $f'_{\rm echo} \simeq (0.00719 ~{\rm sec}) ^{-1}= 139$ Hz for GW170817\footnote{Based on our correspondence with the authors of \cite{Conklin:2017lwb}, it appears that there is a typo in the value of $\Delta t_{\rm echo}$ in the text of the arXiv v.2, while the figures are correct.}, with a p-value of $1/300$. Given the proximity of this value to the second harmonic of our echo frequency $2\times f_{\rm echo} = 144$ Hz, it is plausible that the two different methods are indeed seeing the same echo signal. However, the method applied in  \cite{Conklin:2017lwb} is sub-optimal, as they whiten (rather than Wiener filter) the data (see Sec. \ref{method} above), and thus could underestimate the significance of the correlation peak at $f_{\rm echo} \simeq f'_{\rm echo}/2$. Indeed, our p-value of  $1.6 \times 10^{-5}$ is lower by more than 2 orders of magnitude. Moreover, their $f'_{\rm echo} = 139$ Hz is outside our expected prior range (\ref{f_prior}) for the remnant of the the GW170817 BNS merger, and cannot account for our odd harmonic at $f \simeq 504$ Hz either.  

\end{enumerate}
\section{\label{conclude}Conclusions}

In the past two years, theoretical and observational studies of black hole echoes have emerged as a surprising and yet powerful probe of the quantum nature of black hole horizons. So far, three independent groups have found tentative evidence for echoes in LIGO/Virgo gravitational wave observations of BH (or BNS) mergers at 2-3$\sigma$ level (i.e. p-values of 0.2-2\%, including the look elsewhere effects)  \cite{Conklin:2017lwb,Abedi:2016hgu, Westerweck:2017hus}. In this paper, we presented an optimal and model-agnostic method to search for echo signals well below the ringdown frequency of the black hole quasinormal modes.  As such, this method is well-suited to find echoes in the LIGO/Virgo observations of the BNS merger GW170817. This yielded the first {\it tentative detection} of the echoes at $4.2\sigma$ significance, corresponding to a false detection probability of $1.6 \times 10^{-5}$. The peak signal for echoes occurs 1.0 second after the merger at fundamental frequency of 72 Hz, with another tentative peak at 33 seconds. If confirmed, this could also be the first evidence for formation of a hypermassive neutron star that collapses to form a 2.6-2.7 $M_\odot$ ``black hole'' with dimensionless spin of $0.84-0.87$ within $\sim$ 1 second of the BNS merger. We further estimate that the gravitational wave energy emitted in post-merger echoes is $\sim 3\%$ of the energy emitted in the BNS inspiral, within the LIGO sensitivity band.   

We should note that, unlike the binary BH mergers, there are additional frequencies associated with the accretion disk around BNS merger remnants. While their nature remains mysterious, observed low frequency quasi periodic oscillations (QPOs) in black hole X-ray binaries have similar frequencies to the echoes found here \cite{Motta:2016vwf}\footnote{We thank Avery Broderick for pointing this out.}. Could it be that we are simply detecting the QPOs in (a much more massive) post-merger accretion disk? Or rather, are all low frequency QPO's (of BH accretion disks) sourced by echoes of exotic compact objects?  We shall defer a more detailed analysis of this point to future work, but it suffices to say that the behavior of post BNS merger accretion disks over long time scales are very poorly understood (but see e.g., \cite{Siegel:2017jug} for recent progress in this direction).   

In our last paper \cite{Abedi:2016hgu}, we predicted that: `` ...  a synergy of improvements in observational sensitivity and theoretical modelling can provide conclusive evidence for quantum gravitational alternatives to black hole horizons.'' With the tentative detection reported here, along with other complementary developments, it appears that we are much closer to that possibility. 

{\it Noted Added:} After the publication of this paper on arXiv, Gill, Nathanail, and Rezzolla \cite{Gill:2019bvq} combined Astrophysical arguments with the Electromagnetic observations to find $t_{\rm coll}=0.98^{+0.31}_{-0.26}$ second for formation of a black hole after GW170817 merger event, which is fully consistent with our detection of gravitational wave echoes at $1.0$ second after the merger. 

\acknowledgments
We dedicate this work to the memories of Stephen Hawking and Joe Polchinski, two of the champions of the black hole information paradox, as well as much of  modern theoretical physics, as we know it. Their presence is sorely missed in our community. 
\\

We thank Avery Broderick,  Ofek Birnholtz, Vitor Cardoso, Bob Holdom, Jing Ren, and Daniel Siegel for helpful comments and discussions. We also thank Lam Hui and Shinji Mukohyama for encouraging us to look for echoes in GW170817 BNS merger. We also thank the anonymous referee for helpful comments and suggestions. This work was supported by the University of Waterloo, Natural Sciences and Engineering Research Council of Canada (NSERC), Institute for Research in Fundamental Sciences (IPM), and the Perimeter Institute for Theoretical Physics. Research at the Perimeter Institute is supported by the Government of Canada through Industry Canada, and by the Province of Ontario through the Ministry of Research and Innovation. This research has made use of data, software and/or web tools obtained from the Gravitational Wave Open Science Center (https://www.gw- openscience.org), a service of LIGO Laboratory, the LIGO Scientific Collaboration and the Virgo Collaboration. LIGO is funded by the U.S. National Science Foundation. Virgo is funded by the French Centre National de Recherche Scientifique (CNRS), the Italian Instituto Nazionale della Fisica Nucleare (INFN) and the Dutch Nikhef, with contributions by Polish and Hungarian institutes.


\appendix
\section{Effective time delay between Hanford and Livingston}\label{delta_t}
In order to find the time delay, $\delta t$, between the two detectors, we follow these steps:

\begin{enumerate}
\item      We shift one of the whitened strain series within (-10 msec, 10 msec).
\item      We obtain complex spectrogram for both Hanford and Livingston detectors, $H(t,f)$ and $L(t,f)$ with 16k HZ data.
\item      Given that the two detectors have nearly 180 degrees phase difference, we maximized $|H(t-\delta t,f)-L(t,f)|$ for constant $f$ within the range $t=$(-30 s, +0.5 s) . \item     We only keep the 100 highest values of $$|H(t-\delta t,f)-L(t,f)|_{f=f_{\rm max}(t)},$$ and sum over them (smaller values are more likely to be affected by noise).
\item      Finally, we find the $\delta t$ that maximizes this sum, which happens at the value of $2.62$ ms. 
\end{enumerate}

Note that as spectrogram uses time $\pm 0.5$ sec to Fourier transform around each time, we require to go up to 0.5 sec, to include the merger event. However, there is no overlap between this range used to calculate $\delta t$, and $(0.5 s, 1.5 s)$ range which contributes to our primary echo peak.  

Also note that, while $\delta t = 2.62 $ ms is consistent with the 3 ms time delay reported by LIGO \cite{TheLIGOScientific:2017qsa}, it is not exactly the same as the geometric time delay, as it also captures the small phase difference (away from 180 degrees) between the two detectors.   


\section{Consistency\label{Consistency}}

In this part we support our finding and provide more consistency checks via adding test echoes signal into the detector and investigate its response to our artificial model signal. Then we estimate energy of signal with different methods and provide additional consistency checks with fitting model signal from data. 

\subsection{Injecting artificial echo signal into data}

In order to test whether our method is capable of detecting echoes, we inject a model signal into the region of the data close to BNS merger event, and with same echo time delays (frequency). The main reason behind this is we use the same portion of the data that has similar noise behaviour and detector artifacts. 
 We inject the artificial echo template starting at $t-t_{merger}=-30$ sec and ending at $t-t_{merger}=-20$ sec. We also apply the same phase flip (180 degree) and time shift to the test signal at Hanford and Livingston (for GW170817) as obtained in our paper. Then we use reflection coefficient $\sqrt{R}$ obtained in  \cite{Nakano:2017fvh} in each process of reflection from angular momentum barrier. However, the firewall/fuzzball can change the waveform as well. We shall use $\sqrt{R_{H}} =e^{-\alpha \omega/T_{H}}$  as the reflection coefficient from the``horizon'' which is proposed in \cite{Oshita:2018fqu}. Here $\omega$ is frequency in terms of $\rm{rad}/\rm{sec}$ and $\alpha$ depends on quantum structure of the ``horizon''. We find that $\alpha \simeq 0.015$ produces similar decay times as the observed event. In order to simplify the calculation we assume Schwarzschild approximation of the temperature $T_{H} \sim 1/8\pi M$.  In order to produce the individual echo waveform, we assume initial conditions given by the most persistent quasinormal mode  $\psi(t) \sim e^{-i\omega_{QNM}t} \theta(t)$. 
  
\begin{enumerate}

\item
In frequency domain we have,
\begin{eqnarray}
\psi (\omega)=\frac{1}{\omega - \omega_{QNM}} \label{QNM}
\end{eqnarray}

where the $\omega_{QNM}/2\pi=7641.6-i 863.9\ \rm{Hz}$ represents the (l,m)=(2,2) quasi-normal mode of black hole with mass $2.65 M_{\odot}$ and spin 0.855. This produces a sine function for each echo multiplied by a decaying exponential and a step function at the start time. Then we consider effects coming from initial conditions and superradiance where they might shift the frequencies of echoes to lower frequencies. We postulate a gaussian $e^{-\omega^{2}/\omega_{0}^{2}}$ for these effects multiplied by initial waveform (\ref{QNM}). We set $\omega_{0} = 2\pi\times 1500$ rad/sec.

\item We use reflection coefficient from angular momentum barrier obtained in \cite{Nakano:2017fvh} (Eq. 13) and reflection from quantum structure of the horizon \cite{Oshita:2018fqu}. Amplitude of subsequent n$^{th}$ echo is given as $\left(\sqrt{R_{H}} \times \sqrt{R} \right)^{n}=\left( {e^{-\alpha \omega/T_{H}}R} \right)^{n/2}$. 

\item Time intervals of echoes are set with $\frac{1}{72~ {\rm Hz}}$ = 0.0139 sec.

\item Obtained echo waveforms are injected into the data starting from $t-t_{merger}=-30$ sec and ending at $t-t_{merger}=-20$ sec. 
Obtained template is shown in Fig. \ref{echo_sequence}.
\begin{figure*}[b]
    \includegraphics[width=1\textwidth]{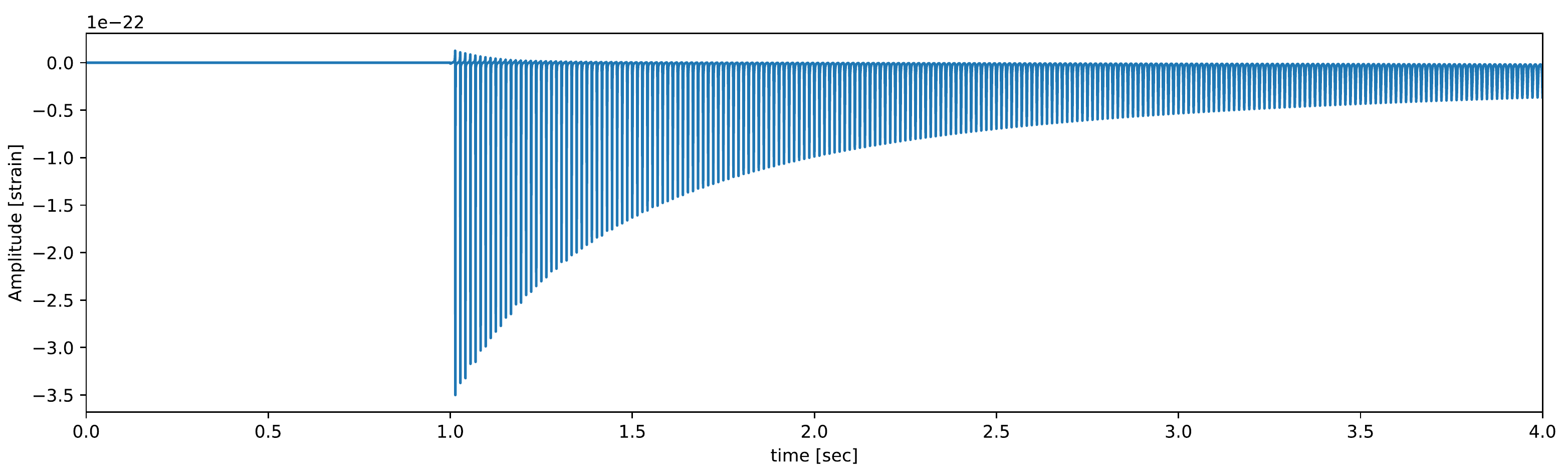}
 \caption{Amplitude-time representations of test model echoes injected into the data with $\alpha=0.015$, $\omega_{0}=2\pi \times 1500$ rad/sec, maximum strain amplitude of $\sim 3.5 \times 10^{-22}$, and time delays of 1/72 sec.}
 \label{echo_sequence}
\end{figure*}

\item We recovered signal of injected echo with maximum strain amplitudes of $\sim (2.0, 2.5, 3.0, 3.5, 4) \times 10^{-22}$ and for $\alpha = 0.015$ and $\omega_{0}=2\pi \times 1500$ rad/sec in Figs. \ref{spectrogram_color_00}, \ref{peak_72Hz_1500}. For smaller values of  $\alpha$ and $\omega_{0}$, the amplitude of detected signal increases. In particular, for the injected maximum amplitude of $3.5 \times 10^{-22}$ [strain], Fig. \ref{peak_72Hz_1500} shows similar amplitude and duration for [$X(t) \sim -6 \times 10^{39}$ and $\Delta t \sim 0.5$ sec] as our observed signal (Fig. 6-7 of the manuscript).  
\begin{figure*}[t]
\centering
    \includegraphics[width=1.0\textwidth]{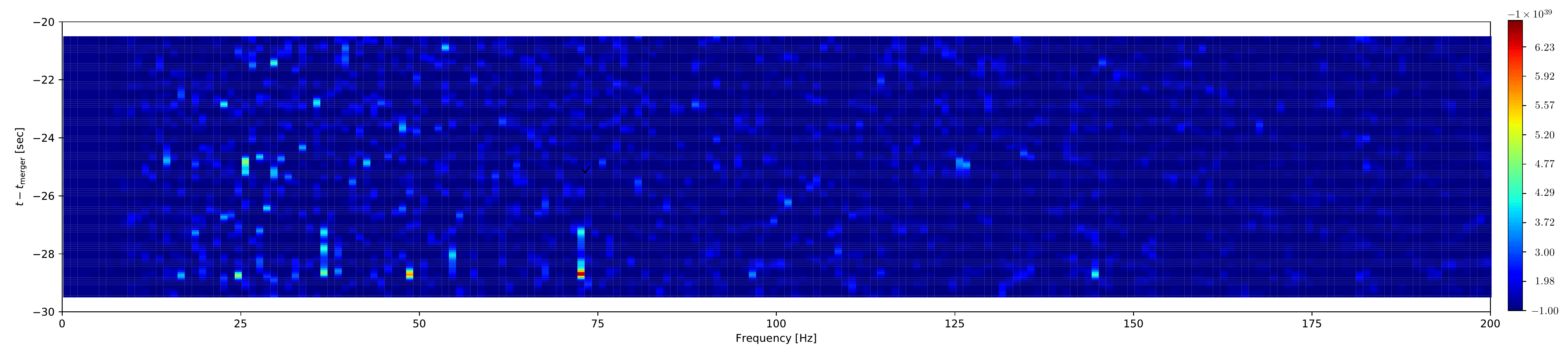}
 \caption{$X(t,f)$ of injected echoes waveform with $\alpha=0.015$, $\omega_{0}=2\pi \times 1500$ rad/sec and maximum echoes amplitude of $3.5 \times 10^{-22}$ [strain] at $t-t_{merger}=(-30,-20)$ sec. The main peak can be seen at 72 Hz and -28.7 sec. }
 \label{spectrogram_color_00}
\end{figure*}

\begin{figure*}[t]
    \includegraphics[width=1\textwidth]{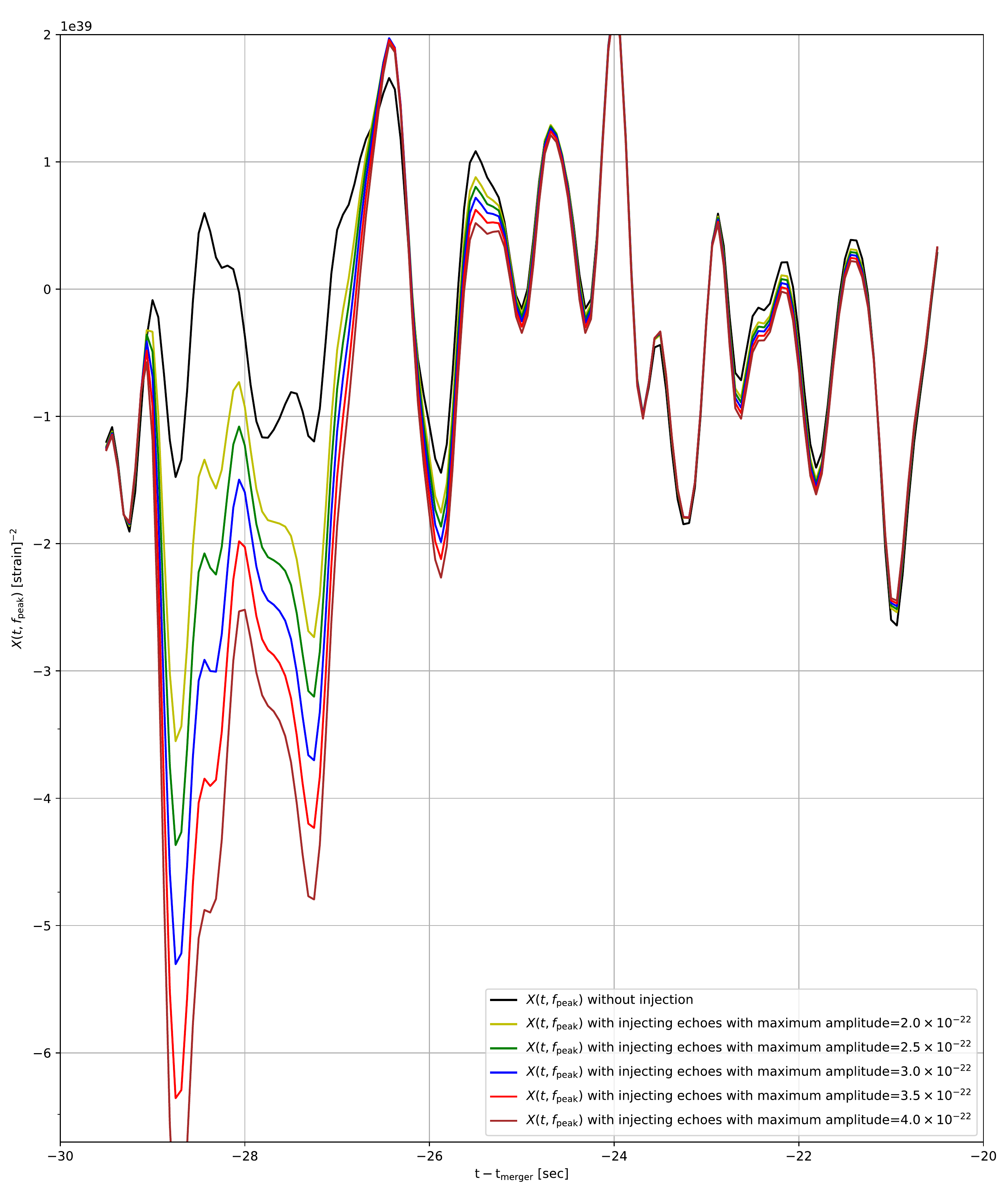}
 \caption{Amplitude-time representations of peaks before the merger $t-t_{merger}=$(-30s,-20s) for injected test echoes signal in Fig. \ref{echo_sequence} with time delays 1/72 sec with amplitudes of $0, 2, 2.5, 3, 3.5, 4 \times 10^{-22}$ strain and $\omega_{0}=2\pi \times 1500$ rad/sec. Echoes with maximum amplitude higher than $3.5\times 10^{-22}$ strain exceed the amplitude of our former detected signal at $-6.48\times10^{39}$.}
 \label{peak_72Hz_1500}
\end{figure*}

\end{enumerate}

\subsubsection{Estimation of energy}

\begin{figure*}[t]
    \includegraphics[width=1\textwidth]{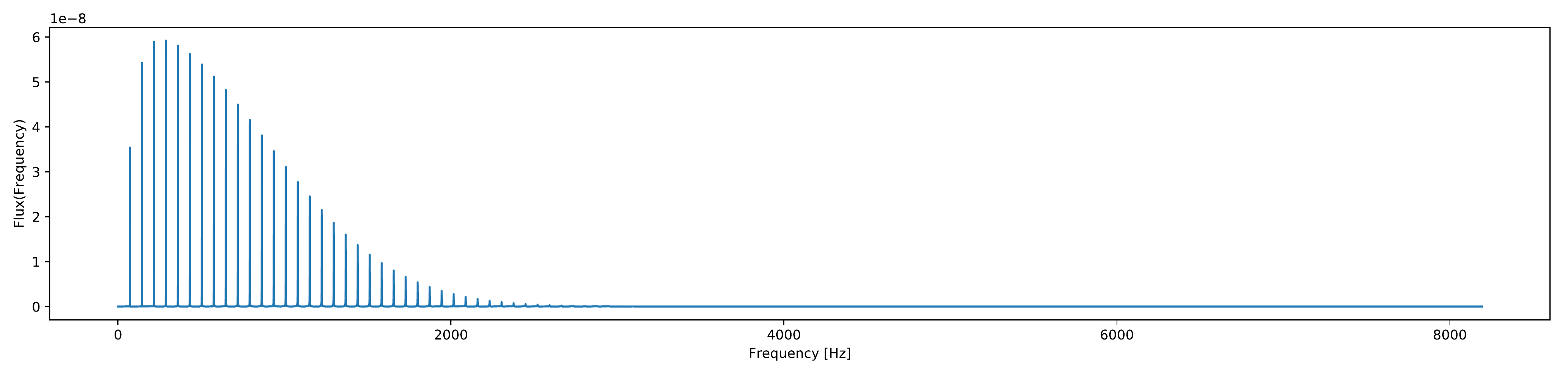}
 \caption{Amplitude-frequency representations of echoes waveform flux in Fig. \ref{echo_sequence} added into the data.}
 \label{Flux}
\end{figure*}

In order to obtain the energy for a given waveform and amplitude we used GW flux (e.g., \cite{Hobson,GravitationalWavePhysics}). For the echo waveform with $\alpha=0.015$, $\omega_{0}=2\pi\times 1500$ rad/sec and maximum amplitude of $\sim 3.5\times 10^{-22}$  (Figures \ref{echo_sequence} and \ref{Flux}) which show similar properties as the actual observed signal (Fig. \ref{peak_72Hz_1500}) energy flux is
\begin{eqnarray}
\!\!\!\!\!\!\!\!\!\!\!\!\!\!\rm{Flux}=\frac{1}{32\pi} \frac{c^{3}}{G} \frac{1}{T} \int_{T}{(\dot{h}_{+}^{2} + \dot{h}_{\times}^{2}})dt=1.0 \times 10^{-5} \frac{\rm{Joul}}{\rm{m}^{2}\rm{s}} \label{eq.3.2}, \nonumber \\
\end{eqnarray}
where Flux is the average energy flow per unit area and unit time within 10 sec of injected echoes. 

Total energy of gravitational wave for this detectable template is given as follows,
\begin{eqnarray}
\rm{Energy} \sim \int_{0}^{\infty} 4\pi r^{2} \times \rm{Flux(t)} dt \ \ \ \ \ \ \ \ \ \ \ \ \ \ \ \ \ \ \nonumber \\
\!\!\!\!\! \sim \int_{0}^{T=10 \rm{sec}} 4\pi r^{2} \times \rm{Flux(t)} dt =0.011 ~M_{\odot}c^2. \label{eq.3.3}
\end{eqnarray}
Here r is event distance which is 40 megaparsecs. 

{\it This estimate is of course very model-dependent}. Taking higher/lower values for $\alpha$ leads to lower/higher energy and higher/lower amplitude for detectable signals respectively. Moreover, the choice of frequency cutoff $\omega_0$ can drastically change energy estimate, as most energy is emitted at higher frequencies. Therefore, the estimate provided in the manuscript (Eq. 5.3)  is consistent with Eq. (\ref{eq.3.3}), even though it is lower by an order of magnitude (= 3\% $\times$ 0.025 $M_{\odot}c^2 \sim 8 \times 10^{-3}$  $M_{\odot}c^2$) as it only includes $f < 500$ Hz, observed by LIGO. 

This suggests that physically motivated echo models with reasonable properties can be detected via our model-agnostic detection method, and exhibit similar characteristics as the observed signal.

\subsubsection{One detector}
If we only inject the above model into one of the detectors, we notice no significant change in $X(t)$ for any physically reasonable amplitude of the injected signal.

\end{document}